\documentstyle[A4,12pt,emlines,epsf]{article}

\title{Quantum Zeno effect in a probed downconversion process}
\author{J. \v{R}eh\'{a}\v{c}ek$^{\#}$\thanks{rehacek@alpha.inf.upol.cz},
J. Pe\v{r}ina$^{\#\$}$,
P. Facchi$^{\mbox{\small \S}}$\thanks{paolo.facchi@ba.infn.it},
S. Pascazio$^{\mbox{\small \S}}$\thanks{saverio.pascazio@ba.infn.it}, and
L. Mi\v{s}ta Jr.$^{\#}$\\
\# \it Department of Optics, Palack\'{y} University,
17. listopadu 50,\\
\it 772~00 Olomouc, Czech Republic\\
\$ \it Joint Laboratory of Optics, Palack\'{y} University and \\
\it Phys. Inst. Czech Acad. Sci.,
17. listopadu 50,\\
\it 772~00 Olomouc, Czech Republic\\
\S \it Dipartimento di Fisica, Universit\`{a} di Bari
and\\
\it  Istituto Nazionale di Fisica Nucleare, Sezione di Bari,\\
\it I-70126 Bari, Italy}
\date{}

\begin{document}
\hfuzz0pt
\maketitle

\begin{abstract}
The distorsion of a spontaneous downconvertion process caused by an
auxiliary mode coupled to the idler wave is analyzed. In general, a
strong coupling with the auxiliary mode tends to hinder the
downconversion in the nonlinear medium. On the other hand, provided
that the evolution is disturbed by the presence of a phase mismatch,
the coupling may increase the speed of downconversion. These effects
are interpreted as being manifestations of quantum Zeno or anti-Zeno
effects, respectively, and they are understood by using the dressed
modes picture of the device. The possibility of using the coupling as
a nontrivial phase--matching technique is pointed out.
\end{abstract}

\section{Introduction}
In quantum optics a downconversion process may be visualized as the
decay of a pump photon into a pair of signal and idler photons of
lower frequency. Provided the pumping is sufficiently strong and
phase matching takes place, the energy of the spontaneously
downconverted light monotonously increases and that of the pump beam
monotonously decreases. From this point of view the downconversion
process may be looked at as the decay process of an unstable system.
It is well known that frequent monitoring of a quantum system leads
to inhibition of its evolution. This phenomenon is called quantum
Zeno effect \cite{others,Misra}. Recently, a thought experiment has
been suggested
\cite{luis1}, in which it is possible to determine the place where
the conversion of the pump photon took place inside the nonlinear
crystal. The idea goes as follows. The nonlinear crystal is
transversely cut in $N$ pieces which are then carefully aligned so
that the signal and pump photons leaving, say, the $k$th slice become
the input signal and pump photons to the $(k+1)$th slice of the
crystal. The idler photons, on the other hand, are removed after each
slice, allowing thus for a future measurement to be performed on
them. If, for example, an ideal detector placed into the path of the
idler mode after the $k$th slice clicks, it is then obvious that the
decay of a pump photon took place somewhere inside the $k$th slice.
By increasing the number of slices, the actual position of birth of
the signal and idler photons becomes more certain. It has been shown
in
\cite{luis1}, in accordance with the Misra-Sudarshan theorem
\cite{Misra}, that the probability of emission of the downconverted
pair decreases with increasing $N$ and for very large number of
crystal slices (continuous observation) the decay of the pump photon
never occurs. It has also been shown \cite{luis2,thun98} that
provided the phase matching condition is not fulfilled in the process
of downconversion, the observation may, on the contrary, {\em
enhance} the emission for a properly chosen $N$ (anti--Zeno or
inverse Zeno effect). This Zeno anti--Zeno interplay has a simple
explanation in terms of destructive and constructive interference of
subsequent emissions inside the nonlinear crystal
\cite{luis1,luis2,thun98}. Here we shall demonstrate that a Zeno-like
behaviour occurs also when instead of cutting the crystal we couple
one of the downconverted beams with an auxiliary mode. Although,
strictly speaking, such a linear coupling cannot be interpreted as
being the realization of a measurement {\em \`a la} von Neumann, the
dynamics of the nonlinear coupler mimics  very well the Zeno
behaviour of the arrangement in \cite{luis1}. It is worth noting, in
this context, that the idea of considering the continuous interaction
with an external agent as a sort of ``steady gaze" at the system goes
back to Kraus \cite{Kraus} and has recently been revived in relation
with the quantum Zeno effect \cite{MPS}. Schulman
\cite{Schulman}, in particular, has even provided a quantitative
relation between the Zeno effect produced by pulsed measurements (in
the sense of
\cite{Misra}) and continuous observation (in the sense discussed
above) performed by an external system.

The paper is organized as follows. In the second section a
theoretical model of the nonlinear coupler is introduced. In the
third section the Zeno--like behavior of the nonlinear coupler is
demonstrated. In the fourth section the dressed modes picture of the
device under investigation is developed and a formal analogy between
a phase mismatch and the coupling of the downconversion process to an
auxiliary mode is explored. Finally, the observed Zeno and anti--Zeno
effects are thoroughly discussed in the fifth section, by using the
obtained results.

\section{Model} \label{sec_model}
\begin{figure}
\centerline{\hspace{-0.5cm}
%TexCad Options
%\grade{\on}
%\emlines{\on}
%\beziermacro{\off}
%\reduce{\on}
%\snapping{\on}
%\quality{2.00}
%\graddiff{0.01}
%\snapasp{1}
%\zoom{1.00}
\special{em:linewidth 0.4pt}
\unitlength 0.70mm
\linethickness{0.4pt}
\begin{picture}(157.00,110.00)
\emline{34.00}{110.00}{1}{125.00}{110.00}{2}
\emline{125.00}{110.00}{3}{125.00}{92.00}{4}
\emline{125.00}{92.00}{5}{34.00}{92.00}{6}
\emline{34.00}{92.00}{7}{34.00}{110.00}{8}
\emline{45.00}{101.00}{9}{46.00}{101.00}{10}
\emline{49.00}{101.00}{11}{51.00}{101.00}{12}
\emline{53.00}{101.00}{13}{58.00}{101.00}{14}
\emline{60.00}{101.00}{15}{65.00}{101.00}{16}
\emline{65.00}{101.00}{17}{66.00}{101.00}{18}
\emline{68.00}{101.00}{19}{81.00}{101.00}{20}
\emline{84.00}{101.00}{21}{101.00}{101.00}{22}
\emline{103.00}{101.00}{23}{125.00}{101.00}{24}
%\vector(128.00,97.00)(137.00,97.00)
\put(137.00,97.00){\vector(1,0){0.2}}
\emline{128.00}{97.00}{25}{137.00}{97.00}{26}
%\end
%\vector(128.00,105.00)(137.00,105.00)
\put(137.00,105.00){\vector(1,0){0.2}}
\emline{128.00}{105.00}{27}{137.00}{105.00}{28}
%\end
\emline{34.00}{83.00}{29}{125.00}{83.00}{30}
\emline{125.00}{83.00}{31}{125.00}{75.00}{32}
\emline{125.00}{75.00}{33}{34.00}{75.00}{34}
\emline{34.00}{75.00}{35}{34.00}{83.00}{36}
%\vector(128.00,79.00)(137.00,79.00)
\put(137.00,79.00){\vector(1,0){0.2}}
\emline{128.00}{79.00}{37}{137.00}{79.00}{38}
%\end
%\vector(40.00,84.00)(40.00,91.00)
\put(40.00,91.00){\vector(0,1){0.2}}
\emline{40.00}{84.00}{39}{40.00}{91.00}{40}
%\end
%\vector(40.00,91.00)(40.00,84.00)
\put(40.00,84.00){\vector(0,-1){0.2}}
\emline{40.00}{91.00}{41}{40.00}{84.00}{42}
%\end
%\vector(55.00,84.00)(55.00,91.00)
\put(55.00,91.00){\vector(0,1){0.2}}
\emline{55.00}{84.00}{43}{55.00}{91.00}{44}
%\end
%\vector(55.00,91.00)(55.00,84.00)
\put(55.00,84.00){\vector(0,-1){0.2}}
\emline{55.00}{91.00}{45}{55.00}{84.00}{46}
%\end
%\vector(70.00,84.00)(70.00,91.00)
\put(70.00,91.00){\vector(0,1){0.2}}
\emline{70.00}{84.00}{47}{70.00}{91.00}{48}
%\end
%\vector(70.00,91.00)(70.00,84.00)
\put(70.00,84.00){\vector(0,-1){0.2}}
\emline{70.00}{91.00}{49}{70.00}{84.00}{50}
%\end
%\vector(85.00,84.00)(85.00,91.00)
\put(85.00,91.00){\vector(0,1){0.2}}
\emline{85.00}{84.00}{51}{85.00}{91.00}{52}
%\end
%\vector(85.00,91.00)(85.00,84.00)
\put(85.00,84.00){\vector(0,-1){0.2}}
\emline{85.00}{91.00}{53}{85.00}{84.00}{54}
%\end
%\vector(100.00,84.00)(100.00,91.00)
\put(100.00,91.00){\vector(0,1){0.2}}
\emline{100.00}{84.00}{55}{100.00}{91.00}{56}
%\end
%\vector(100.00,91.00)(100.00,84.00)
\put(100.00,84.00){\vector(0,-1){0.2}}
\emline{100.00}{91.00}{57}{100.00}{84.00}{58}
%\end
%\vector(115.00,84.00)(115.00,91.00)
\put(115.00,91.00){\vector(0,1){0.2}}
\emline{115.00}{84.00}{59}{115.00}{91.00}{60}
%\end
%\vector(115.00,91.00)(115.00,84.00)
\put(115.00,84.00){\vector(0,-1){0.2}}
\emline{115.00}{91.00}{61}{115.00}{84.00}{62}
%\end
\put(40.00,101.00){\makebox(0,0)[cc]{$\bf \Gamma$}}
\put(133.00,109.00){\makebox(0,0)[cc]{{\scriptsize\bf signal}}}
\put(133.00,92.00){\makebox(0,0)[cc]{{\scriptsize \bf idler}}}
\put(132.50,83.00){\makebox(0,0)[cc]{{\scriptsize \bf aux}}}
\put(47.00,87.00){\makebox(0,0)[cc]{$\bf \kappa$}}
\emline{19.00}{102.00}{63}{27.00}{102.00}{64}
\emline{19.00}{100.00}{65}{27.00}{100.00}{66}
\emline{31.00}{101.00}{67}{25.00}{104.00}{68}
\emline{31.00}{101.00}{69}{25.00}{98.00}{70}
\put(23.00,107.00){\makebox(0,0)[cc]{{\scriptsize\bf pump}}}
\put(78.00,79.00){\makebox(0,0)[cc]{\small linear waveguide}}
\put(80.00,96.00){\makebox(0,0)[cc]{\small nonlinear waveguide}}
\end{picture}}
\vspace{-5 cm}
\caption{Outline of the nonlinear coupler}\label{Fig_outline}
\end{figure}
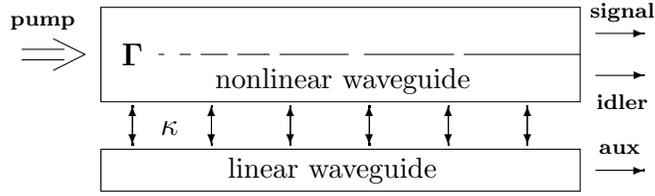

Consider a nonlinear coupler made up of two waveguides, through which
four modes, pump $p$, signal $s$, idler $i$, and auxiliary mode $b$
propagate in the same direction, see Fig.~\ref{Fig_outline}. The
nonlinear waveguide is filled with a second-order nonlinear medium in
which ultra--violet pump photons are downconverted to signal and
idler photons of lower frequency. In addition, the idler mode is
allowed to exchange energy, e.g. by means of evanescent waves, with
the auxiliary mode $b$ propagating through a linear medium.

In the
following we will assume that all four modes are monochromatic and
their frequencies are fixed, e.g. by placing narrow interference
filters in front of detectors. Provided the amplitudes of the fields
inside the coupler vary little during an optical period (SVEA
approximation), and provided the linear coupling is sufficiently weak
so that it can be described by coupled modes theory (Born
approximation) \cite{coupl}, the effective Hamiltonian of our device
reads ($\hbar=1$)
%%%%%%%%%%%%%%%%%%%%%%%%%%%%%%%%%%%%%%%%%%%%%%%%%%%%%%%%%%%%%%%%%%%%%
\begin{equation}\label{hamilt_all}
H=\omega_p a_p^{\dagger}a_p+\omega_s a_s^{\dagger}a_s+
\omega_i a_i^{\dagger}a_i+
\omega_b b^{\dagger}b+\left(\Gamma a_p a_s^{\dagger} a_i^{\dagger}
e^{i\Delta t}+\kappa a_i^{\dagger}b+\mbox{h.c.}\right).
\end{equation}
%%%%%%%%%%%%%%%%%%%%%%%%%%%%%%%%%%%%%%%%%%%%%%%%%%%%%%%%%%%%%%%%%%%%%
Here $\omega_{\alpha}$ is the frequency of mode $\alpha$,
$\Delta$=$({\bf k}_{p}-{\bf k}_{s}-{\bf k}_{i})_z$ is the nonlinear
phase mismatch, $\Gamma$ and $\kappa$ are the nonlinear and linear
coupling constants, respectively, and the propagation variable $z$
has been replaced with the evolution parameter $t$. Usually, $\kappa$
is proportional to the overlap between the idler and auxiliary modes
\cite{coupl}, whereas the nonlinear coupling constant $\Gamma$ is
proportional to the second order nonlinear susceptibility $\chi(2)$
\cite{hong85}. It is convenient to split the Hamiltonian
(\ref{hamilt_all}) into free and interaction parts
%%%%%%%%%%%%%%%%%%%%%%%%%%%%%%%%%%%%%%%%%%%%%%%%%%%%%%%%%%%%%%%%%%%%%
\begin{equation}\label{perturb}
H=H_0+H_I.
\end{equation}
%%%%%%%%%%%%%%%%%%%%%%%%%%%%%%%%%%%%%%%%%%%%%%%%%%%%%%%%%%%%%%%%%%%%%
In order to get rid of the free evolution in the Heisenberg equations
of motion
%%%%%%%%%%%%%%%%%%%%%%%%%%%%%%%%%%%%%%%%%%%%%%%%%%%%%%%%%%%%%%%%%%%%%
\begin{equation}\label{heisenberg}
\dot{a}=-i [a,H_0+H_I],
\end{equation}
%%%%%%%%%%%%%%%%%%%%%%%%%%%%%%%%%%%%%%%%%%%%%%%%%%%%%%%%%%%%%%%%%%%%%
where $a$ is the annihilation operator of a particular mode, we
introduce the new field operators
%%%%%%%%%%%%%%%%%%%%%%%%%%%%%%%%%%%%%%%%%%%%%%%%%%%%%%%%%%%%%%%%%%%%%
\begin{equation}\label{free_evol}
a'_\alpha=e^{i\omega_\alpha t}a_\alpha , \qquad (\alpha=p,s,i)
\end{equation}
%%%%%%%%%%%%%%%%%%%%%%%%%%%%%%%%%%%%%%%%%%%%%%%%%%%%%%%%%%%%%%%%%%%%%
and analogously for $b$. Substituting these new variables together
with the Hamiltonian (\ref{perturb}) into Eq.~(\ref{heisenberg}), we
arrive at the equations of motion
%%%%%%%%%%%%%%%%%%%%%%%%%%%%%%%%%%%%%%%%%%%%%%%%%%%%%%%%%%%%%%%%%%%%%
\begin{equation}\label{new_eq}
\dot{a}'=-i [a',H_I'],
\end{equation}
where
%%%%%%%%%%%%%%%%%%%%%%%%%%%%%%%%%%%%%%%%%%%%%%%%%%%%%%%%%%%%%%%%%%%%%
\begin{equation}\label{new_hamilt}
H_I'=\Gamma a'_p a_s^{\prime\dagger}a_i^{\prime\dagger} e^{i\Delta
t}e^{-i(\omega_p-\omega_s-\omega_i)t} +
\kappa a_i^{\prime\dagger}b' e^{i(\omega_i-\omega_b)t} + \mbox{h.c.}.
\end{equation}
%%%%%%%%%%%%%%%%%%%%%%%%%%%%%%%%%%%%%%%%%%%%%%%%%%%%%%%%%%%%%%%%%%%%%
Because the Hamiltonian ({\ref{hamilt_all}) contains products of
three operators, the equations of motion (\ref{heisenberg}) and
(\ref{new_eq}) are nonlinear. The nonlinearity accounts mainly for
saturation effects and must be taken into account whenever the pump
beam becomes depleted (e.g.\ medium in a cavity). On the other hand,
if the pumping is sufficiently strong and if the nonlinear
interaction is weak so that only a small fraction of the pump photons
is removed from the input beam, we can simplify our problem by
describing the strong pump wave in classical terms, i.e. we let
$a_p$=$\xi\exp(i\omega_pt)$, where $\xi$ and $\omega_p$ denote the
complex amplitude and the frequency of the classical pump wave,
respectively. With the help of the strong pump wave approximation the
interaction Hamiltonian of our problem (\ref{new_hamilt}) is
simplified as follows
%%%%%%%%%%%%%%%%%%%%%%%%%%%%%%%%%%%%%%%%%%%%%%%%%%%%%%%%%%%%%%%%%%%%%
\begin{equation}\label{dc_hamilt}
H_I=\Gamma a_s^{\dagger}a_i^{\dagger}e^{i\Delta t}+
\kappa a_i^{\dagger}b + \mbox{h.c.},
\end{equation}
%%%%%%%%%%%%%%%%%%%%%%%%%%%%%%%%%%%%%%%%%%%%%%%%%%%%%%%%%%%%%%%%%%%%%
where we assumed that the frequency matching conditions hold:
$\omega_p-\omega_s-\omega_i=0$ and $\omega_b=\omega_i$. The amplitude
$\xi$ has been absorbed in coupling constant $\Gamma$ and all
operators are written without apostrophes, for simplicity. The
dynamics of the nonlinear coupler (\ref{dc_hamilt}) reduces to the
dynamics of the phase matched spontaneous downconversion process
provided that $\kappa$=$\Delta$=$0$ and the initial state is taken as
$|\Psi_0\rangle$=$|vac\rangle_s\otimes|vac\rangle_i$. As we already
mentioned in the introduction, the average number of signal and idler
photons originating in the crystal of length $L$,
%%%%%%%%%%%%%%%%%%%%%%%%%%%%%%%%%%%%%%%%%%%%%%%%%%%%%%%%%%%%%%%%%%%
\begin{equation}\label{simple}
\langle a_{s,i}^{\dagger}a_{s,i}
\rangle_{\mbox{\protect{\scriptsize vac}}}=\sinh^2\Gamma L,
\qquad (\kappa=\Delta=0)
\end{equation}
%%%%%%%%%%%%%%%%%%%%%%%%%%%%%%%%%%%%%%%%%%%%%%%%%%%%%%%%%%%%%%%%%%%%
is then an (exponentially) increasing function of $L$.

\section{Linear coupling turned on} \label{sec_matched}
The behaviour of the downconversion process dramatically changes when
one of the two downconverted modes (e.g.\ the idler mode) is coupled
to an auxiliary mode via a linear interaction. The Hamiltonian
(\ref{dc_hamilt}) yields, when $\Delta=0$ (phase matching),
%%%%%%%%%%%%%%%%%%%%%%%%%%%%%%%%%%%%%%%%%%%%%%%%%%%%%%%%%%%%%%%%%%%%%
\begin{eqnarray}\label{mis_motion}
\dot{a}_s&=&-i\Gamma a_i^{\dagger},\nonumber\\
\dot{a}_i&=&-i\Gamma a_s^{\dagger}-i\kappa b,
\qquad (\Delta=0) \nonumber\\
\dot{b}&=&-i\kappa a_i
\end{eqnarray}
%%%%%%%%%%%%%%%%%%%%%%%%%%%%%%%%%%%%%%%%%%%%%%%%%%%%%%%%%%%%%%%%%%%%%
and we are interested in the regime of weak nonlinearity, expressed
by the condition $\kappa>\Gamma$. Notice that two opposite tendencies
compete in Eqs.\ (\ref{mis_motion}): an elliptic structure, leading
to oscillatory behavior, governed by the coupling parameter $\kappa$,
\begin{equation}\label{elliptic}
\ddot{a}_i = -\kappa^2 a_i ,\qquad\qquad
\ddot{b} = -\kappa^2 b
\end{equation}
and a hyperbolic structure, yielding exponential behavior, governed
by the nonlinear parameter $\Gamma$,
\begin{equation}\label{hyperbolic}
\ddot{a}_s = \Gamma^2 a_s ,\qquad\qquad
\ddot{a}_i = \Gamma^2 a_i .
\end{equation}
The threshold between these two regimes occurs for $\Gamma=\kappa$.

The system of equations (\ref{mis_motion}) is easily solved and the
number of output signal photons, which is the same as the number of
pump photons decays, reads
%%%%%%%%%%%%%%%%%%%%%%%%%%%%%%%%%%%%%%%%%%%%%%%%%%%%%%%%%%%%%%%%%%%%%
\begin{equation}\label{mis_vac}
\langle a_s^{\dagger} a_s\rangle_{\mbox{\protect{\scriptsize vac}}}
=\frac{\Gamma^2}{\chi^2}\sin^2\chi L+
\frac{\kappa^2\Gamma^2}{\chi^4}(1-\cos\chi L)^2,
\end{equation}
%%%%%%%%%%%%%%%%%%%%%%%%%%%%%%%%%%%%%%%%%%%%%%%%%%%%%%%%%%%%%%%%%%%%%
where $\chi$=$\sqrt{\kappa^2-\Gamma^2}$. Hereafter, the symbol
$\langle\dots\rangle_{\mbox{\protect{\scriptsize vac}}}$ denotes
averaging with respect to the initial vacuum state
$|\Psi_0\rangle$=$|vac\rangle_s\otimes|vac\rangle_i\otimes|vac\rangle_b$
\cite{init_state}. Unlike the case of phase matched
downconversion (\ref{simple}), the exchange of energy between all
modes now becomes periodical when $\kappa>\Gamma$. As the linear
coupling becomes stronger, the period of the oscillations gets
shorter and the amplitude of the oscillations decreases as
$\kappa^{-2}$, namely
%%%%%%%%%%%%%%%%%%%%%%%%%%%%%%%%%%%%%%%%%%%%%%%%%%%%%%%%%%%%%%%%%%%%%
\begin{equation}\label{mis_vac_asympt}
\langle a_s^{\dagger} a_s\rangle_{\mbox{\protect{\scriptsize vac}}}
\sim\frac{\Gamma^2}{\kappa^2}\sin^2\kappa L+
\frac{\Gamma^2}{\kappa^2}(1-\cos\kappa L)^2
=\frac{4\Gamma^2}{\kappa^2}\sin^2\frac{\kappa L}{2}
\qquad (\kappa\gg\Gamma).
\end{equation}
%%%%%%%%%%%%%%%%%%%%%%%%%%%%%%%%%%%%%%%%%%%%%%%%%%%%%%%%%%%%%%%%%%%%%
For very strong coupling \cite{large_kappa} the downconversion
process is completely frozen, the medium becomes effectively linear
and the pump photons propagate throught it without ``decay." Notice
that in this situation, even if $L$ is increased, the number of
downconverted photons is bounded [compare with the opposite case
(\ref{simple})]. This can be interpreted as a manifestation of
quantum Zeno effect in the following sense: by increasing the
coupling with the auxiliary mode, one performs a better
``observation" of the idler mode and therefore of the ``decay" of the
pump. The hindering of the evolution results. There is an intuitive
explanation of this behavior: since the linear coupling changes the
phases of the amplitudes of the interacting modes, the constructive
interference yielding exponential increase of the converted energy
(\ref{simple}) is destroyed, and downconversion becomes frozen. We
shall come back to this point and corroborate this intuitive picture
in the next section.

The proposed interpretation in terms of quantum Zeno effect is
readily understandable and rather appealing. On the other hand, one
should remark that since only the output fields are accessible to
measurement in the experimental setup in Fig.~\ref{Fig_outline}, no
relevant information is readily available about the place where the
signal and idler photon are created \cite{observable}. In this sense,
no {\em bona fide} measurement is being performed on the fields. The
situation would be different if we provided the auxiliary waveguide
with some photodetection device like an array of highly efficient
photodetectors. For sufficiently strong linear coupling, the decay
product (the idler photon) would enter the auxiliary mode soon after
the emission, it could then be detected by a pixel of the
photodetection array and we could thereafter infer the place where
the emission had taken place. As there is no such a detection device
present in the setup in question, the {\em coherent} superposition of
the two possibilities: ``the idler photon is in the idler mode'' and
``the idler photon is in the auxiliary mode'', is maintained through
the evolution and no decomposition of the wave function occurs.
Nevertheless, it is still possible (and useful) to speak about
quantum Zeno effect in the more general sense given above. A
discussion of this point is given in
\cite{nakazato96} in connection with the experiment performed by
Itano {\em et al.} \cite{itano90}.

\section{Dressed modes}  \label{sec_dressed}
We now look for the modes dressed by the interaction $\kappa$. This
will provide an alternative interpretation and a more rigorous
explanation of the result obtained above. Let us diagonalize the
Hamiltonian (\ref{hamilt_all}) with respect to the linear coupling.
By setting $\omega_i=\omega_b$ and $\kappa$ real, it is easy to see
that in terms of the dressed modes
%%%%%%%%%%%%%%%%%%%%%%%%%%%%%%%%%%%%%%%%%%%%%%%%%%%%%%%%%%%%%%%%%%%%%
\begin{eqnarray}\label{dressed_modes}
c &=& (a_i+b)/\sqrt{2}, \nonumber\\ d &=& (a_i-b)/\sqrt{2},
\end{eqnarray}
%%%%%%%%%%%%%%%%%%%%%%%%%%%%%%%%%%%%%%%%%%%%%%%%%%%%%%%%%%%%%%%%%%%%%
the Hamiltonian (\ref{hamilt_all}) reads
%%%%%%%%%%%%%%%%%%%%%%%%%%%%%%%%%%%%%%%%%%%%%%%%%%%%%%%%%%%%%%%%%%%%%
\begin{eqnarray}\label{hamilt_all_dress}
H &=& \omega_p a_p^{\dagger}a_p+\omega_s a_s^{\dagger}a_s+
\omega_c c^{\dagger}c+
\omega_d d^{\dagger}d\nonumber\\
& & +
\frac{\Gamma}{\sqrt{2}} a_p a_s^{\dagger}c^{\dagger}e^{i\Delta t} +
\frac{\Gamma}{\sqrt{2}} a_p a_s^{\dagger}d^{\dagger}e^{i\Delta t}
+ \mbox{h.c.},
\end{eqnarray}
%%%%%%%%%%%%%%%%%%%%%%%%%%%%%%%%%%%%%%%%%%%%%%%%%%%%%%%%%%%%%%%%%%%%%
where the dressed energies are
%%%%%%%%%%%%%%%%%%%%%%%%%%%%%%%%%%%%%%%%%%%%%%%%%%%%%%%%%%%%%%%%%%%%%
\begin{eqnarray}\label{dressed_energies}
\omega_c &=& \omega_i + \kappa, \nonumber\\
\omega_d &=& \omega_i - \kappa.
\end{eqnarray}
%%%%%%%%%%%%%%%%%%%%%%%%%%%%%%%%%%%%%%%%%%%%%%%%%%%%%%%%%%%%%%%%%%%%%
If $\Delta=0$, in the strong pump limit, by following the same
procedure of section
\ref{sec_model}, instead of (\ref{dc_hamilt}), we get the following
interaction Hamiltonian
%%%%%%%%%%%%%%%%%%%%%%%%%%%%%%%%%%%%%%%%%%%%%%%%%%%%%%%%%%%%%%%%%%%%%
\begin{equation}\label{dc_hamilt_dress}
H_I=\frac{\Gamma}{\sqrt{2}} a_s^{\dagger}c^{\dagger}e^{i\kappa t}+
\frac{\Gamma}{\sqrt{2}} a_s^{\dagger}d^{\dagger}e^{-i\kappa t}
+ \mbox{h.c.}, \qquad (\Delta=0)
\end{equation}
%%%%%%%%%%%%%%%%%%%%%%%%%%%%%%%%%%%%%%%%%%%%%%%%%%%%%%%%%%%%%%%%%%%%%
where we assumed as before that the frequency matching conditions
holds: $\omega_p-\omega_s-\omega_i=0$. By comparing the Hamiltonian
(\ref{dc_hamilt_dress}) with the Hamiltonian (\ref{dc_hamilt}) when
$\kappa=0$:
%%%%%%%%%%%%%%%%%%%%%%%%%%%%%%%%%%%%%%%%%%%%%%%%%%%%%%%%%%%%%%%%%%%%%
\begin{equation}\label{dc_hamilt_k0}
H_I=\Gamma a_s^{\dagger}a_i^{\dagger}e^{i\Delta t}+
\mbox{h.c.}, \qquad (\kappa=0)
\end{equation}
%%%%%%%%%%%%%%%%%%%%%%%%%%%%%%%%%%%%%%%%%%%%%%%%%%%%%%%%%%%%%%%%%%%%%
describing downconversion with phase mismatch $\Delta$, it is
apparent that the coupling and the phase mismatch influence the
downconversion process in the same way. In fact for large values of
the phase mismatch $\Delta$ it is easy to find that
%%%%%%%%%%%%%%%%%%%%%%%%%%%%%%%%%%%%%%%%%%%%%%%%%%%%%%%%%%%%%%%%%%%%%
\begin{equation}\label{mis_vac_asympt_Delta}
\langle a_s^{\dagger} a_s\rangle_{\mbox{\protect{\scriptsize vac}}}
\sim\frac{4\Gamma^2}{\Delta^2}\sin^2\frac{\Delta L}{2}
\qquad (\Delta\gg\Gamma),
\end{equation}
%%%%%%%%%%%%%%%%%%%%%%%%%%%%%%%%%%%%%%%%%%%%%%%%%%%%%%%%%%%%%%%%%%%%%
which is to be compared with Eq.~(\ref{mis_vac_asympt}). The coupling
of the idler mode $a_i$ with the auxiliary mode $b$ yields two
dressed modes $c$ and $d$ the pump photon can decay to. They are
completely decoupled and due to their energy shift
(\ref{dressed_energies}), exhibit a phase mismatch $\pm\kappa$. Since
the phase mismatch effectively shortens the time during which a fixed
phase relation holds between the interacting beams, the amount of
converted energy is smaller than in the ideal case of perfectly phase
matched interaction. This explains the results of
section~\ref{sec_matched}. A strong linear coupling then makes the
subsequent emissions of converted photons interfere destructively and
the nonlinear interaction is frozen. In this respect the disturbances
caused by the coupling and by frequently repeated measurements are
similar and we can interpret the phenomenon as a quantum Zeno effect.

\section{Competition between the coupling and the mismatch} \label{sec_competition}
In the previous section we saw that the nonlinear interaction was
affected by both linear coupling and phase mismatch in the same way.
Namely, the effectiveness of the nonlinear process dropped down under
their action. In this section we show that when both disturbing
elements are present in the dynamics of the downconversion process,
the linear coupling can, rather surprisingly, compensate for the
phase mismatch and vice versa, so that the probability of emission of
the signal and idler photons can almost return back to its
undisturbed value.

We start from the equations of motion generated by the full
interaction Hamiltonian (\ref{dc_hamilt})
%%%%%%%%%%%%%%%%%%%%%%%%%%%%%%%%%%%%%%%%%%%%%%%%%%%%%%%%%%%%%%%%%%%%%
\begin{eqnarray}\label{full_eq}
\dot{a}_s&=&-i\Gamma a_i^{\dagger}e^{i\Delta t},\nonumber\\
\dot{a}_i&=&-i\Gamma a_s^{\dagger}e^{i\Delta t}-
i\kappa b, \qquad (\Delta \neq 0, \kappa \neq 0)\nonumber\\
\dot{b}&=&-i\kappa a_i.
\end{eqnarray}
%%%%%%%%%%%%%%%%%%%%%%%%%%%%%%%%%%%%%%%%%%%%%%%%%%%%%%%%%%%%%%%%%%%%%
Although it is easy to write down the explicit solution of the system
(\ref{full_eq}), we shall here provide only a qualitative discussion
of the solution. The main features are then best demonstrated with
the help of a few figures. Eliminating idler and auxiliary mode
variables from Eq.~(\ref{full_eq}) we get a differential equation of
the third order for the annihilation operator of the signal mode. Its
characteristic polynomial (upon substitution
$a_s(t)$=$a_s(0)\exp(i\lambda t)$)
%%%%%%%%%%%%%%%%%%%%%%%%%%%%%%%%%%%%%%%%%%%%%%%%%%%%%%%%%%%%%%%%%%%%%
\begin{equation}\label{polynom}
\lambda^3+2\Delta\lambda^2+(\Delta^2-\kappa^2+\Gamma^2)\lambda
+\Delta\Gamma^2, \quad \kappa\neq 0
\end{equation}
%%%%%%%%%%%%%%%%%%%%%%%%%%%%%%%%%%%%%%%%%%%%%%%%%%%%%%%%%%%%%%%%%%%%%
is recognized as a cubic polynomial in $\lambda$ with real
coefficients. An oscillatory behaviour of the signal mode occurs only
provided the polynomial (\ref{polynom}) has three real roots (causus
irreducibilis), i.e.\ its determinant $D$ must obey the condition
$D<0$. Expanding the determinant in the small nonlinear coupling
parameter $\Gamma$ and keeping terms up to the second order in
$\Gamma$ we obtain
%%%%%%%%%%%%%%%%%%%%%%%%%%%%%%%%%%%%%%%%%%%%%%%%%%%%%%%%%%%%%%%%%%%%%
\begin{equation}\label{determ}
D=-\frac{\kappa^2}{27}\left[(\kappa^2-\Delta^2)^2-
(5\Delta^2+3\kappa^2)\Gamma^2\right], \quad \Gamma\ll \Delta,\kappa.
\end{equation}
%%%%%%%%%%%%%%%%%%%%%%%%%%%%%%%%%%%%%%%%%%%%%%%%%%%%%%%%%%%%%%%%%%%%%
It is seen that a mismatched downconversion behaves in either
oscillatory or hyperbolic way, depending on the strength of the
coupling with the auxiliary mode. The values of $\kappa$ lying at the
boundary between these two types of dynamics are determined by
solving the equation $D=0$. The only two nontrivial solutions are
%%%%%%%%%%%%%%%%%%%%%%%%%%%%%%%%%%%%%%%%%%%%%%%%%%%%%%%%%%%%%%%%%%%%%
\begin{equation}\label{boundary}
\kappa_{1,2}=\sqrt{\Delta^2+\frac{3}{2}\Gamma^2\pm\sqrt{8}\Delta
\Gamma}.
\end{equation}
%%%%%%%%%%%%%%%%%%%%%%%%%%%%%%%%%%%%%%%%%%%%%%%%%%%%%%%%%%%%%%%%%%%%%
The case $\Delta\gg\Gamma$ is of main interest in this section
(otherwise we have the situation already described in
section~\ref{sec_matched}). Hence we can, eventually, drop $\Gamma^2$
in Eq.~(\ref{boundary}). The resulting intervals are
%%%%%%%%%%%%%%%%%%%%%%%%%%%%%%%%%%%%%%%%%%%%%%%%%%%%%%%%%%%%%%%%%%%%%
\begin{equation} \label{boundary_approx}
\begin{array}{lll}
\mbox{hyperbolic behaviour:}&\quad&
\kappa\in\langle\Delta-\sqrt{2}\Gamma,\Delta+\sqrt{2}\Gamma\rangle\\
\mbox{oscillatory behaviour:}&\quad&
\kappa\in\langle 0,\Delta-\sqrt{2}\Gamma) \cup
(\Delta+\sqrt{2}\Gamma,\infty).
\end{array}
\end{equation}
%%%%%%%%%%%%%%%%%%%%%%%%%%%%%%%%%%%%%%%%%%%%%%%%%%%%%%%%%%%%%%%%%%%%%

\begin{figure}
\vspace{-7cm}
\hspace*{3.3cm}
\epsfxsize=11cm \epsfbox{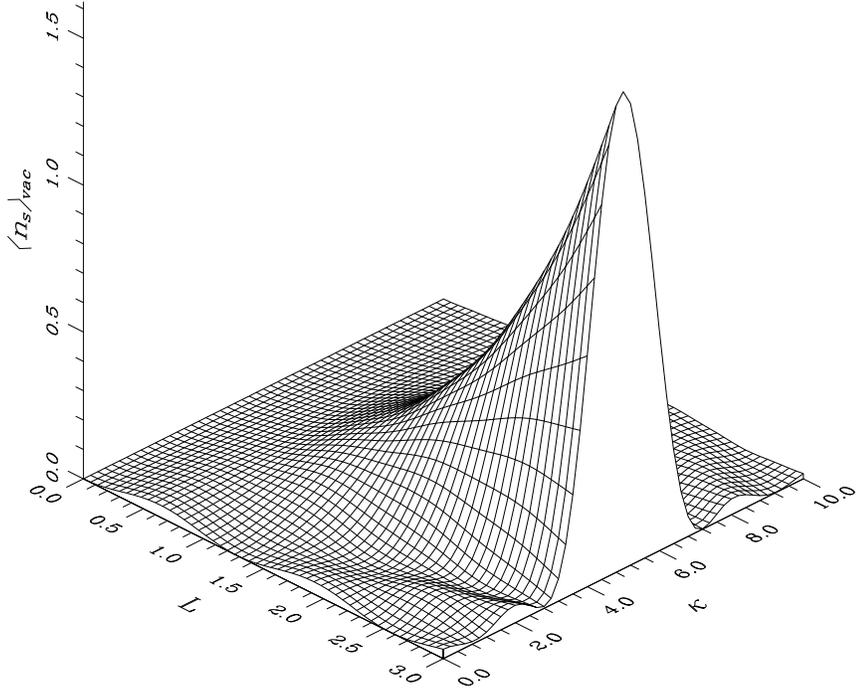}
\vspace{0.5cm}
\caption{Mean number of signal photons
$\langle n_s\rangle$ behind the nonlinear medium as a function of
interaction length $L$  and strength $\kappa$ of linear coupling. The
nonlinear mismatch and nonlinear coupling parameter are $\Delta$=$5$
and $\Gamma$=$0.5$, respectively.}
\label{Fig_intens_vs_kappa}
\vspace{-0.2cm}
\end{figure}
\begin{figure}
\vspace*{-8.6cm}
\hspace*{3.3cm}
\epsfxsize=11cm \epsfbox{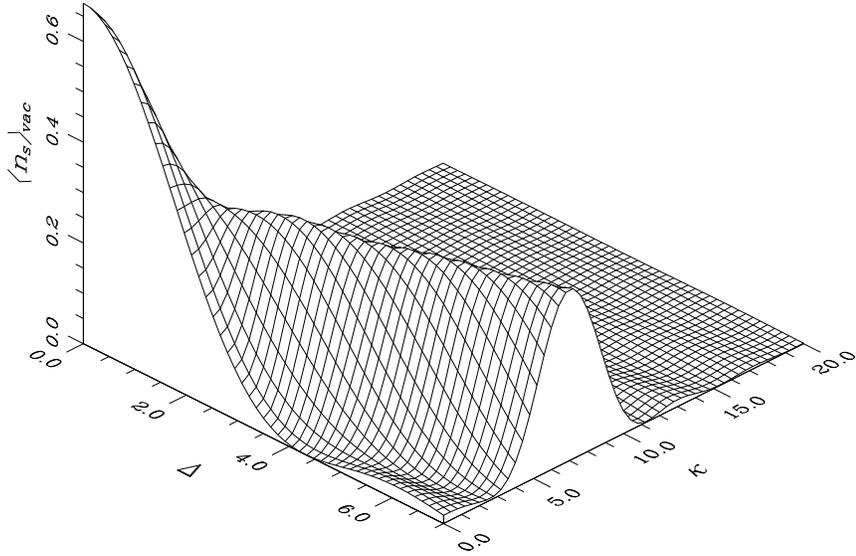}
\vspace{0.5cm}
\caption{Interplay between linear coupling and
phase mismatch. The mean number of signal photons $\langle
n_s\rangle$ behind the nonlinear medium of length $L$=$1.5$ is shown
vs strength $\kappa$ of linear coupling  and nonlinear mismatch
$\Delta$. The nonlinear coupling parameter is fixed at
$\Gamma$=$0.5$.}\label{Fig_delta_vs_kappa}
\vspace{0cm}
\end{figure}

The behaviour of the mismatched downconversion process is shown in
Fig.~\ref{Fig_intens_vs_kappa} for a particular choice of $\Delta$.
In absence of linear coupling the downconverted light shows
oscillations and the overall effectiveness of the nonlinear process
is small due to the presence of phase mismatch $\Delta$. However, as
we switch on the coupling between the idler and auxiliary mode, the
situation changes. By increasing the strength of the coupling the
period of the oscillations gets longer and its amplitude gets larger.
When $\kappa$ becomes larger than $\Delta-\sqrt{2}\Gamma$ the
oscillations are no longer seen and the intensity of the signal beam
starts to grow monotonously. We can say that in this regime the
initial nonlinear mismatch has been compensated by the coupling.

The interplay between nonlinear mismatch and linear coupling is
illustrated in Fig.~\ref{Fig_delta_vs_kappa}. A significant
production of signal photons is a clear manifestation of an
anti--Zeno effect. In correspondence with the observations in
\cite{luis2,thun98}, such an anti--Zeno effect occurs only provided
a substantial phase mismatch is introduced in the process of
downconversion. It is worthwile to compare the interesting behavior
seen in Fig.~\ref{Fig_delta_vs_kappa} with the Zeno and anti--Zeno
effects observed in a sliced nonlinear crystal (Fig.~1 in
\cite{thun98}). It can be seen that the coupling parameter $\kappa$
here plays a role similar to the number of slices $N$, into which the
crystal is cut in the latter scheme. Moreover, the sharpness of the
``observation'' ($\kappa$ or $N$), at which a maximum output
intensity occurs, is approximately a linear function of the
introduced phase mismatch in both schemes. There are, however, also
some points of difference. For example, the maximum output intensity
obtainable for a given $\Delta$ by slicing the crystal decreases with
increasing phase mismatch $\Delta$ \cite{thun98}. On the other hand,
no matter how strong the mismatch is, it can always be removed with
the help of a suitable linear coupling (and {\em vice versa}). This
difference is due to the $1/N$ scaling of intensities of output light
generated by a process under observation
\cite{luis1,luis2,thun98}. An analogous factor is missing here,
in Eq.~(\ref{mis_vac}).

Several intuitive explanations of the anti--Zeno like behaviour seen
in Fig.~\ref{Fig_delta_vs_kappa} are at hand. From the point of view
of constructive and destructive interference one can say that since
the linear coupling effectively changes the phase relations among
interacting modes, the destructive interference of subsequent pump
photon decays caused by phase mismatch is suppressed in the same way
as the constructive interference has been suppressed in the case of
perfectly matched interaction.

Fig.~\ref{Fig_delta_vs_kappa} can also be interpreted in a
quantitative way in analogy with the dressed state description of
interaction of atoms with intense light \cite{pascazio99}. In terms
of the dressed modes $c$ and $d$ of Eq.~(\ref{dressed_modes}), if
$\Delta\neq 0$, in place of the Hamiltonian (\ref{dc_hamilt_dress})
one gets
%%%%%%%%%%%%%%%%%%%%%%%%%%%%%%%%%%%%%%%%%%%%%%%%%%%%%%%%%%%%%%%%%%%%%
\begin{equation}\label{dc_hamilt_dress1}
H_I=\frac{\Gamma}{\sqrt{2}}
a_s^{\dagger}c^{\dagger}e^{i(\Delta+\kappa) t}+
\frac{\Gamma}{\sqrt{2}} a_s^{\dagger}d^{\dagger}e^{i(\Delta-\kappa) t}
+ \mbox{h.c.},
\end{equation}
%%%%%%%%%%%%%%%%%%%%%%%%%%%%%%%%%%%%%%%%%%%%%%%%%%%%%%%%%%%%%%%%%%%%%
that yields the equations of motion
%%%%%%%%%%%%%%%%%%%%%%%%%%%%%%%%%%%%%%%%%%%%%%%%%%%%%%%%%%%%%%%%%%%%%
\begin{eqnarray}\label{full_eq_dress}
\dot{a}_s&=&-i\frac{\Gamma}{\sqrt{2}}
c^{\dagger}e^{i(\Delta+\kappa) t}-i\frac{\Gamma}{\sqrt{2}}
d^{\dagger}e^{i(\Delta-\kappa) t},\nonumber\\
\dot{c}&=&-i\frac{\Gamma}{\sqrt{2}}
a_s^{\dagger}e^{i(\Delta+\kappa) t},\nonumber\\
\dot{d}&=&-i\frac{\Gamma}{\sqrt{2}}
a_s^{\dagger}e^{i(\Delta-\kappa) t}.
\end{eqnarray}
%%%%%%%%%%%%%%%%%%%%%%%%%%%%%%%%%%%%%%%%%%%%%%%%%%%%%%%%%%%%%%%%%%%%%
The energy scheme implied by Eq.~(\ref{full_eq_dress}) is shown in
Fig.~\ref{Fig_energy_scheme}. Under the influence of the coupling
with the auxiliary mode $b$ the mismatched downconversion splits into
two dressed energy--shifted interactions. It is apparent that when
$\kappa=\pm\Delta$ one of the two interactions becomes resonant. The
other one is  ``counterrotating" and acquires a phase mismatch
$2\Delta$, yielding oscillations. Also, the amplitude of such
oscillations decreases as $\Delta^{-2}$ and the mode output  becomes
negligible compared to the other one. The use of the rotating wave
approximation in Eq.~(\ref{full_eq_dress}) is fully justified in this
case and the system is easily solved. The output signal intensity
reads

\begin{figure}
\vspace{0.5cm}
\centerline{\hspace{-3.5cm}
%TexCad Options
%\grade{\on}
%\emlines{\on}
%\beziermacro{\off}
%\reduce{\on}
%\snapping{\on}
%\quality{2.00}
%\graddiff{0.01}
%\snapasp{1}
%\zoom{1.00}
\special{em:linewidth 0.4pt}
\unitlength 1.00mm
\linethickness{0.4pt}
\begin{picture}(107.00,92.00)
\emline{25.00}{60.00}{1}{60.00}{60.00}{2}
\emline{25.00}{77.00}{3}{28.00}{77.00}{4}
\emline{29.00}{77.00}{5}{32.00}{77.00}{6}
\emline{33.00}{77.00}{7}{36.00}{77.00}{8}
\emline{37.00}{77.00}{9}{40.00}{77.00}{10}
\emline{41.00}{77.00}{11}{44.00}{77.00}{12}
\emline{45.00}{77.00}{13}{48.00}{77.00}{14}
\emline{49.00}{77.00}{15}{52.00}{77.00}{16}
\emline{53.00}{77.00}{17}{56.00}{77.00}{18}
\emline{57.00}{77.00}{19}{60.00}{77.00}{20}
\emline{60.00}{77.00}{21}{64.00}{92.00}{22}
\emline{64.00}{92.00}{23}{70.00}{92.00}{24}
\emline{60.00}{77.00}{25}{64.00}{62.00}{26}
\emline{64.00}{62.00}{27}{70.00}{62.00}{28}
\emline{76.00}{62.00}{29}{79.00}{62.00}{30}
\emline{80.00}{62.00}{31}{83.00}{62.00}{32}
\emline{84.00}{62.00}{33}{87.00}{62.00}{34}
\emline{88.00}{62.00}{35}{91.00}{62.00}{36}
\emline{92.00}{62.00}{37}{95.00}{62.00}{38}
\emline{96.00}{62.00}{39}{99.00}{62.00}{40}
\emline{100.00}{62.00}{41}{103.00}{62.00}{42}
\emline{104.00}{62.00}{43}{107.00}{62.00}{44}
\emline{76.00}{92.00}{45}{79.00}{92.00}{46}
\emline{80.00}{92.00}{47}{83.00}{92.00}{48}
\emline{84.00}{92.00}{49}{87.00}{92.00}{50}
\emline{88.00}{92.00}{51}{91.00}{92.00}{52}
\emline{92.00}{92.00}{53}{95.00}{92.00}{54}
\emline{96.00}{92.00}{55}{99.00}{92.00}{56}
\emline{100.00}{92.00}{57}{103.00}{92.00}{58}
\emline{104.00}{92.00}{59}{107.00}{92.00}{60}
\emline{76.00}{60.00}{61}{107.00}{60.00}{62}
\emline{70.00}{62.00}{63}{72.00}{62.00}{64}
\emline{70.00}{92.00}{65}{72.00}{92.00}{66}
%\vector(40.00,60.00)(40.00,77.00)
\put(40.00,77.00){\vector(0,1){0.2}}
\emline{40.00}{60.00}{67}{40.00}{77.00}{68}
%\end
%\vector(40.00,77.00)(40.00,60.00)
\put(40.00,60.00){\vector(0,-1){0.2}}
\emline{40.00}{77.00}{69}{40.00}{60.00}{70}
%\end
%\vector(82.00,60.00)(82.00,92.00)
\put(82.00,92.00){\vector(0,1){0.2}}
\emline{82.00}{60.00}{71}{82.00}{92.00}{72}
%\end
%\vector(82.00,92.00)(82.00,60.00)
\put(82.00,60.00){\vector(0,-1){0.2}}
\emline{82.00}{92.00}{73}{82.00}{60.00}{74}
%\end
%\vector(101.00,57.00)(101.00,60.00)
\put(101.00,60.00){\vector(0,1){0.2}}
\emline{101.00}{57.00}{75}{101.00}{60.00}{76}
%\end
%\vector(101.00,65.00)(101.00,62.00)
\put(101.00,62.00){\vector(0,-1){0.2}}
\emline{101.00}{65.00}{77}{101.00}{62.00}{78}
%\end
\put(44.00,68.00){\makebox(0,0)[cc]{$\Delta$}}
\put(86.00,81.00){\makebox(0,0)[lc]{$\Delta+\kappa$}}
\put(104.00,66.00){\makebox(0,0)[lc]{$\Delta-\kappa$}}
\end{picture}}
\vspace{-5.5 cm}
\caption{Energy scheme of a mismatched downconversion
process subject to linear coupling. The bottom solid lines denote a
resonant process.}\label{Fig_energy_scheme}
\end{figure}

%%%%%%%%%%%%%%%%%%%%%%%%%%%%%%%%%%%%%%%%%%%%%%%%%%%%%%%%%
\begin{equation}\label{qpm}
\langle a_s^{\dagger}
a_s\rangle_{\mbox{\protect{\scriptsize vac}}}
=\sinh^2\left(\frac{\Gamma}{\sqrt{2}}L\right),
\quad (\kappa=\pm\Delta) \quad (\Delta\gg 1/L)
\end{equation}
%%%%%%%%%%%%%%%%%%%%%%%%%%%%%%%%%%%%%%%%%%%%%%%%%%%%%%%%%
[compare with Eq.\ (\ref{simple})]. The linear coupling to an
auxiliary mode compensates for the phase mismatch up to a change in
the effective nonlinear coupling strength
$\Gamma\rightarrow\Gamma/\sqrt{2}$.
% It follows from Eq.~(\ref{qpm})
%that an anti--Zeno enhancement of probability of pump photon
%``decay'' up to  $50\%$ of its ``undisturbed" value (\ref{simple})
%is possible, if the interaction length is short.

As a matter of fact, the condition $\kappa=\pm\Delta$ can be
interpreted also as a condition for achieving the so--called
quasi--phase--matching in the nonlinear process. A
quasi--phase--matched regime of generation
\cite{qpm} is usually
forced by creating an artificial lattice inside a nonlinear medium,
e.g. by periodic modulation of the nonlinear coupling coefficient.
Periodic change of sign of $\Gamma$ (rectangular modulation) yields
the effective coupling strength $\Gamma\rightarrow 2\Gamma/\pi$
\cite{qpm}, where, as before, $\Gamma$ is the coupling strength of
the phase--matched interaction. Thus the continuous ``observation''
of the idler mode even gives a slightly better enhancement of the
decay rate than the most common quasi--phase--matching technique.

To summarize, the statement ``the downconversion process is
mismatched'' means that the nonlinear process is out of resonance in
the sense that the momentum of the decay products (signal and idler
photons) differs from the momentum carried by the pump photon before
the decay took place. When the linear interaction is switched on the
system gets dressed and the energy spectrum changes. A careful
adjustment of the coupling strength $\kappa$ makes then possible to
tune the nonlinear interaction back to resonance. In this way the
probability of pump photon decay can be greatly enhanced. This occurs
when $\kappa\simeq \pm\Delta$ and explains why the anti--Zeno effect
takes place along the line $\kappa = \Delta$ in
Fig.~\ref{Fig_delta_vs_kappa}.

\section{Conclusion}
In this article a downconversion process disturbed by the presence of
a linear coupling between the idler and some auxiliary mode has been
discussed. Although, strictly speaking, such a coupling is not a
measurement in von Neumann's sense, we found a striking similarity
between the dynamics of our system and the dynamics of the
downconversion processes taking place in a sliced nonlinear crystal,
where a Zeno interpretation is feasible and appealing.

In some sense, the Zeno effect is a consequence of the new dynamical
features introduced by the coupling with an external agent that
(through its interaction) ``looks closely" at the system. When this
interaction can be effectively described as a projection operator
{\em \`a la} von Neumann, we obtain the usual formulation of the
quantum Zeno effect in the limit of very frequent measurements. In
general, the description in terms of projection operators may not
apply, but the dynamics can be modified in a way that is strongly
reminiscent of Zeno. Examples of the type analyzed in this paper call
for a broader definition of ``quantum Zeno effect."

\vspace*{10pt}
\noindent
{\large \bf Acknowledgments} We thank Dr. A. Luis for some pertinent
remarks. We acknowledge support by grant No VS96028, by research
project CEZ:J14/98 ``Wave and particle optics'' of the Czech Ministry
of Education and by the TMR-Network ERB-FMRX-CT96-0057 of the
European Union.


\begin{thebibliography}{9}
\bibitem{init_state} Other choices of the initial state are
possible as well. Different input states of mode $b$ then correspond
to different input states of the pointer of a measuring apparatus.
\bibitem{large_kappa} In the regime of very large $\kappa$, however,
the coupled modes theory breaks down and some other experimental
realization of the Hamiltonian (\ref{hamilt_all}) should be found.
\end{thebibliography}
\end{document}